\documentclass[preprint2]{emulateapj}
\usepackage{psfig}
\catcode`\@=11
\def\gsim{\ifmmode{\,\mathrel{\mathpalette\@versim>\,}}
    \else{$\,\mathrel{\mathpalette\@versim>}\,$}\fi}
\def\lsim{\ifmmode{\,\mathrel{\mathpalette\@versim<\,}}
    \else{$\,\mathrel{\mathpalette\@versim<}\,$}\fi}
\def\@versim#1#2{\lower 2.9truept \vbox{\baselineskip 0pt \lineskip
    0.5truept \ialign{$\m@th#1\hfil##\hfil$\crcr#2\crcr\sim\crcr}}}
\catcode`\@=12  
\def\km{{\rm \,km}}

\def\sminus{{\rm \,s^{-1}}}

\def\kpc{{\rm \,kpc}}

\def\Mstarprime{M_*^{\prime}}
\def\cstarprime{c_*^{\prime}}
\def\Reprime{\Re^{\prime}}
\def\sgetprime{\sget^{\prime}}
\def\Mstar{M_*}

\def\Mstartilde{\tilde{M}_*}

\def\MBH{M_{\rm BH}}
\def\Ns{N_{\rm s}}

\def\Msun{M_{\odot}}

\def\Mlens{M_{\rm lens}}

\def\Re{R_{\rm e}}

\def\rstartilde{\tilde{r}_{*}}
\def\fstar{f_*}

\def\sigmazero{\sigma_0}

\def\sget{\sigma_{\rm e2}}
\def\sgetsq{\sigma_{\rm e2}^2}

\def\cstar{c_*}

\def\betar{\beta_{R}}
\def\betac{\beta_{c}}
\def\betas{\beta_{\sigma}}

\slugcomment{Accepted, 14 October 2009}

\shorttitle{Dry merging and size evolution of galaxies}
\shortauthors{Nipoti et al.}

\begin{document}

%% LaTeX will automatically break titles if they run longer than
%% one line. However, you may use \\ to force a line break if
%% you desire.

\title{Can dry merging explain the size evolution of early-type galaxies?}

%% Use \author, \affil, and the \and command to format
%% author and affiliation information.
%% Note that \email has replaced the old \authoremail command
%% from AASTeX v4.0. You can use \email to mark an email address
%% anywhere in the paper, not just in the front matter.
%% As in the title, use \\ to force line breaks.

\author{C.~Nipoti}
\affil{Dipartimento di Astronomia,
Universit\`a di Bologna, via Ranzani 1, I-40127 Bologna, Italy}
\email{carlo.nipoti@unibo.it}
\author{T.~Treu and M.~W.~Auger}
\affil{Department of Physics, University of California, Santa Barbara, CA
93106-9530, USA}
\and
\author{A.~S.~Bolton}
\affil{Institute for Astronomy, University of Hawaii, 2680 Woodlawn Dr., Honolulu, HI 96822 USA}

%% Notice that each of these authors has alternate affiliations, which
%% are identified by the \altaffilmark after each name.  Specify alternate
%% affiliation information with \altaffiltext, with one command per each
%% affiliation.

%\altaffiltext{1}{Visiting Astronomer, Cerro Tololo Inter-American Observatory.
%CTIO is operated by AURA, Inc.\ under contract to the National Science
%Foundation.}

%% Mark off your abstract in the ``abstract'' environment. In the manuscript
%% style, abstract will output a Received/Accepted line after the
%% title and affiliation information. No date will appear since the author
%% does not have this information. The dates will be filled in by the
%% editorial office after submission.

\begin{abstract}
  The characteristic size of early-type galaxies (ETGs) of given
  stellar mass is observed to increase significantly with cosmic time,
  from redshift $z\gsim 2$ to the present. A popular explanation for
  this size evolution is that ETGs grow through dissipationless
  (``dry'') mergers, thus becoming less compact. Combining N-body
  simulations with up-to-date scaling relations of local ETGs, we show
  that such an explanation is problematic, because dry mergers do not
  decrease the galaxy stellar-mass surface-density enough to explain
  the observed size evolution, and also introduce substantial scatter
  in the scaling relations. Based on our set of simulations, we
  estimate that major and minor dry mergers increase half-light radius
  and projected velocity dispersion with stellar mass as $\Re\propto
  \Mstar^{1.09\pm0.29}$ and $\sget\propto \Mstar^{0.07\pm0.11}$,
  respectively. This implies that: 1) if the high-$z$ ETGs are indeed as
  dense as estimated, they cannot evolve into present-day ETGs via dry
  mergers; 2) present-day ETGs cannot have assembled more than $\sim
  45\%$ of their stellar mass via dry mergers. Alternatively, dry
  mergers could be reconciled with the observations if there was
  extreme fine tuning between merger history and galaxy properties, at
  variance with our assumptions.  Full cosmological simulations will
  be needed to evaluate whether this fine-tuned solution is
  acceptable.
\end{abstract}

%% Keywords should appear after the \end{abstract} command. The uncommented
%% example has been keyed in ApJ style. See the instructions to authors
%% for the journal to which you are submitting your paper to determine
%% what keyword punctuation is appropriate.


%% From the front matter, we move on to the body of the paper.
%% In the first two sections, notice the use of the natbib \citep
%% and \citet commands to identify citations.  The citations are
%% tied to the reference list via symbolic KEYs. The KEY corresponds
%% to the KEY in the \bibitem in the reference list below. We have
%% chosen the first three characters of the first author's name plus
%% the last two numeral of the year of publication as our KEY for
%% each reference.

\section{Introduction}

In the hierarchical model of structure formation, galaxy mergers are
expected to play a central role in the assembly and evolution of
galaxies. If present-day early-type galaxies (ETGs) have experienced
mergers in relatively recent times ($z\lsim 1-2$), most of these
mergers must have been dissipationless or ``dry'', because the old
stellar populations of ETGs \citep[e.g.][]{Thomas05} are inconsistent
with substantial recent star formation.

Present-day ETGs satisfy several empirical correlations between
half-light radius $\Re$, central stellar velocity dispersion
$\sigmazero$, bulge luminosity $L$, stellar mass $\Mstar$, black-hole
mass $\MBH$ and gravitational lensing mass $\Mlens$, such as the
$L$-$\sigmazero$ \citep{Fab76}, $L$-$\Re$ \citep{Kor77}, $\MBH$-$L$
\citep{Mag98}, $\MBH$-$\sigmazero$ \citep{Ferrarese2000,Geb00},
$\MBH$-$\Mstar$ \citep{Marconi03}, $\Mlens$-$\Re$ and
$\Mlens$-$\sigmazero$ relations~\citep[][hereafter NTB09]{Nip09},
%but also correlations between three quantities, such as 
the Fundamental Plane \citep[hereafter FP, relating $L$, $\Re$ and
$\sigmazero$;][]{Dre87,Djo87} and the lensing Mass Plane
\citep[relating $\Mlens$, $\sigmazero$ and
$\Re$;][]{Bol07,Bol08b}. The existence of these scaling laws has been
used to establish that present-day ETGs did not form by several dry
mergers of lower mass progenitors {\it obeying the same scaling laws},
because as an effect of dry mergers the galaxy size increases rapidly
with mass, while the velocity dispersion remains almost constant, in
contrast with the observed
correlations~\citep[][NTB09]{Ciotti01,Nip03,Boy06,Ciotti07}.
 
Photometric and spectroscopic observations of high-redshift ($z\gsim
1-2$) ETGs suggest that these objects may be remarkably more compact
than their local counterparts \citep[e.g.][hereafter vD08; Saracco,
Longhetti, \& Andreon
2009]{stiav99,Dad05,Tru06,Zir07,Cimatti08,vdW08,vDo08}.  These
findings contributed to renew interest in the proposal that dry
merging might be a key process in the growth of ETGs, because it is
one of the few known mechanisms able to make galaxies less compact
\citep[e.g.][]{Kho06,Hop09a,Naab09,vdW09}.  However, although dry
merging works qualitatively in the right direction, it is not clear
whether it works quantitatively.

In this Letter we address the question of the viability of dry merging
as a process driving the structural evolution of ETGs, by testing
whether it is able to produce the local scaling laws with their
relatively small intrinsic scatters.  For this purpose, we use the set
of numerical simulations of dissipationless minor and major mergers
presented in NTB09 and test them against the stellar-mass scaling
relations of the Sloan Lenses ACS Survey (SLACS) sample of lens ETGs
\citep{Auger09}. This strategy allows us to construct a direct link
between the observed properties of high-redshift ($z\gsim1-2$)
galaxies---for which estimates of the stellar masses are
available---and realistic simulated galaxies, in which the relative
distribution of luminous and dark matter has been calibrated on the
lensing-mass scaling relations (NTB09).

Throughout the paper we adopt \cite{Sal55} Initial Mass Function
(IMF), which is consistent with the gravitational lensing constraints
for SLACS ETGs \citep{Tre09b}. Observational data are converted to
this calibration when necessary. The main results of this paper do not
depend on this choice, since they involve stellar-mass ratios and the
IMF normalization effectively factors out.

\section{Observed stellar-mass scaling relations}
\label{sec:obs}

In the present work we consider the stellar-mass--size,
stellar-mass--velocity-dispersion and stellar-mass FP relations
derived by \cite{Auger09} for the SLACS sample of lens ETGs, assuming
Salpeter IMF stellar masses and using the velocity dispersions and
rest-frame V-band half-light radii derived from Sloan Digital Sky
Survey (SDSS) spectroscopy and Hubble Space Telescope photometry.  The
correlations are obtained using galaxies with stellar masses generally
in the range $11\lsim \log \Mstar/\Msun \lsim 12$.

The stellar-mass--size relationship is given by
\begin{eqnarray}
\nonumber &\log& \left({\Re \over \kpc}\right) = (0.730 \pm 0.047) \\
&\times& \log \left({\Mstar\over 10^{11}\Msun}\right) + 0.391 \pm 0.029,
\label{eq:reff}
\end{eqnarray}
with intrinsic scatter in $\log \Re$ at fixed mass 0.071. The
stellar-mass--velocity-dispersion relation is
\begin{eqnarray}
\nonumber &\log& \left({\sget \over \km\sminus}\right) = (0.196 \pm 0.033) \\
&\times& \log \left({\Mstar\over 10^{11}\Msun}\right) + 2.281 \pm 0.021,
\label{eq:sigma}
\end{eqnarray}
where $\sget$ is the projected velocity dispersion within an aperture
radius $\Re/2$. This has an intrinsic scatter in $\log \sget$ at fixed
mass of 0.055. Finally we consider the FP-like correlation
\begin{eqnarray}
\nonumber &\log&\,\cstar = (0.099 \pm 0.067) \\
&\times& \log \left({\Mstar\over 10^{11}\Msun}\right) -0.959 \pm 0.042,
\label{eq:mdim}
\end{eqnarray}
where $\cstar \equiv \sgetsq\Re/2G\Mstar$ is a dimensionless structure
parameter.
%the ratio between the the
%so-called ``dimensional mass'' \citep{Bol08b} and the stellar mass.
The intrinsic scatter in $\log \cstar$ at fixed $\Mstar$ is 0.077.
%It must be noted that the slope $(0.099 \pm 0.067)$ in
%equation~(\ref{eq:mdim}) is inconsistent (within 1-$\sigma$) with
%zero, meaning that---at least over the mass range probed by the SLACS
%sample---the stellar-mass FP has significant tilt
%\citep[see][]{Auger09}, in contrast with the lensing Mass Plane
%\citep{Bol07,Bol08b}.

We recall that SLACS lenses are indistinguishable from control samples
of SDSS galaxies with the same stellar velocity dispersion and size,
in terms of luminosity
%surface brightness, location on the FP, 
and distribution of environments \citep{Bol06,Tre06,Bol08a,Tre09a}. In
addition, as discussed in \cite{Auger09}, the above scaling relations
are consistent with other scaling relations derived for larger samples
of SDSS galaxies with a similar range of masses
\citep[e.g.][]{Hyde09}: in particular the intrinsic scatter of the
$\Mstar$-$\Re$ relation is consistent with the observed scatter of the
widely used correlation found by \citet{Shen03}.

%%%%%%%%%%%%%%%
%%%% FIG 1
%%%%%%%%%%%%%%%%
\begin{figure}
%\epsscale{1.0}
%\plotone{hze_sm.eps}
\centerline{
\psfig{figure=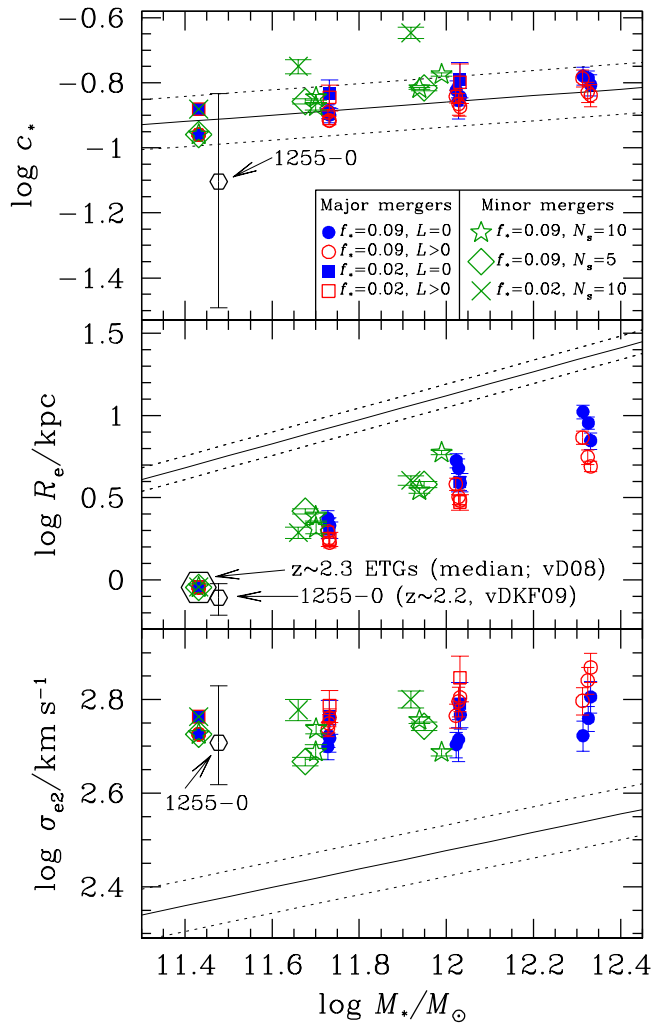,width=\hsize,angle=0,bbllx=40bp,bblly=695bp,bburx=370bp,bbury=150bp,clip=}}
\caption{Structural parameter $\cstar\equiv\sgetsq\Re/2G\Mstar$) (top
  panel), effective radius (central panel) and projected velocity
  dispersion (bottom panel) as functions of the stellar mass.  The
  solid and dashed lines represent the local observed correlations
  (equations~\ref{eq:reff}-\ref{eq:mdim}), with the associated
  intrinsic scatters.  The big hexagon represents the median $\Mstar$
  and $\Re$ for a sample of nine $z\sim2.3$ ETGs (vD08). The small
  hexagons with error bars represent the ETG 1255-0 ($z=2.186$;
  vDKF09). The other points and error bars represent the average
  values and the 1-$\sigma$ scatters (due to projection effects) for
  the simulated galaxies of the merger hierarchies indicated in the
  top panel ($\fstar$ is the stellar mass fraction, $L$ is the orbital
  angular momentum modulus and $\Ns$ is the number of satellites per
  step).}
\label{fig:sm}
\end{figure}

%%%%%%%%%%%%%%%
%%%% FIG 2
%%%%%%%%%%%%%%%%
\begin{figure}
%\epsscale{1.0}
%\plotone{hze_scatt.eps}
\centerline{ 
\psfig{figure=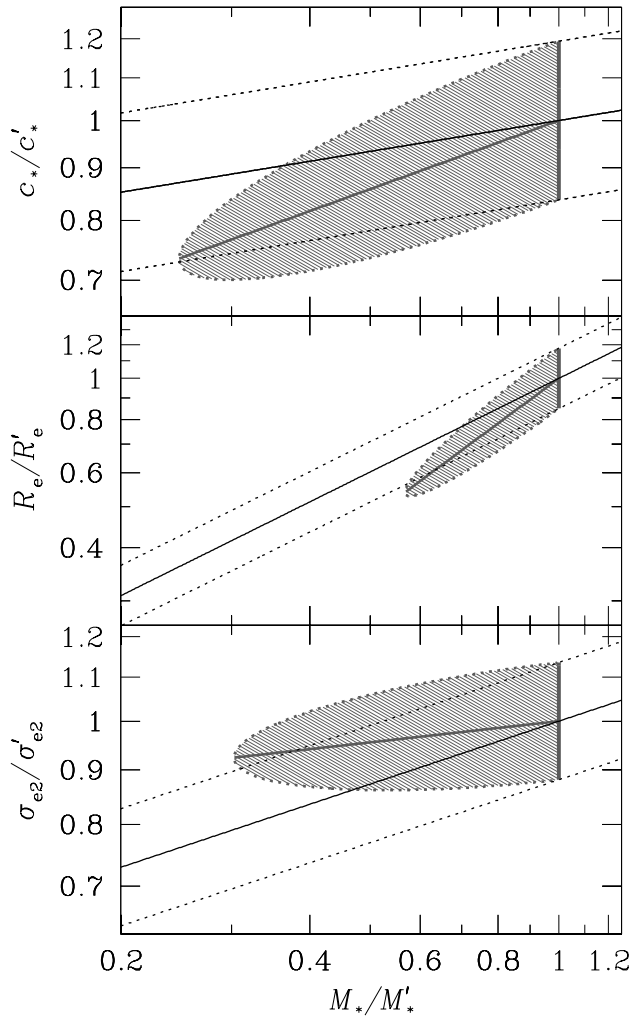,width=\hsize,angle=0,bbllx=40bp,bblly=695bp,bburx=370bp,bbury=150bp,clip=}} 
\caption{The shaded areas represent the distributions in the
  $\Mstar$-$\cstar$, $\Mstar$-$\Re$ and $\Mstar$-$\sget$ planes of
  allowed dry-merging progenitors of local ETGs with stellar mass
  $\Mstarprime$ (vertical bars).  The upper and lower boundaries of
  the shaded areas correspond to 1-$\sigma$ of the distributions,
  while the solid lines within the shaded areas show their average
  values. $\cstarprime$, $\Reprime$ and $\sgetprime$ are the average
  values of $\cstar$, $\Re$, and $\sget$ for local ETGs with mass
  $\Mstarprime$.  The local observed correlations are plotted as in
  Fig.~\ref{fig:sm}.}
\label{fig:scatt}
\end{figure}

\section{Effect of dry mergers on compact high-$z$ early-type galaxies}
\label{sec:highz}

Here we combine the scaling laws described above, observations of
high-$z$ ETGs and the results of the N-body simulations presented in
NTB09 to address the question of whether dry mergers can be
responsible for the observed size evolution of ETGs. The set of
simulations (see NTB09 for details) consists of both major and minor
dry-merger hierarchies, i.e. sequences of two or three steps in each
of which the galaxy mass increases by a factor of $\sim 2$ as a
consequence of either a binary equal-mass merger or the accretion of
$\Ns$ smaller satellites.  The seed galaxies and the satellites are
spherical, with \citet{Nav96} dark-matter distribution, \citet{Deh93}
stellar distribution (with central logarithmic slope $\gamma=1.5$),
and no gas. Altogether we consider eight major-merger hierarchies
(differing in the orbital energy and angular momentum of the galaxy
encounters, and in the stellar mass fraction $\fstar$ of the seed
galaxies) and four minor-merger hierarchies (differing in $\fstar$,
$\Ns$ and in the phase-space distribution of the satellites' centers
of mass).  Each galaxy collision is followed up to the virialization
of the merger remnant. In the second and third steps of the
hierarchies the remnants of the previous steps are used to realize the
initial conditions.

The seed galaxies (and consequently all the end-products of each
merging hierarchy) can be rescaled by fixing arbitrarily the
simulation mass and length scales $\Mstartilde$ and $\rstartilde$ (see
NTB09). Though the relative effect of dry mergers on the galaxy
structure and kinematics is independent of $\Mstartilde$ and
$\rstartilde$, it is useful to fix these quantities, without loss of
generality, to make the comparison with observations
straightforward. For our purpose, it is natural to consider seed
galaxies with stellar masses and sizes similar to those of high-$z$
ETGs. Here we take as reference the sample of nine $z\sim2.3$ ETGs
studied by vD08, having median stellar mass
$\langle\Mstar\rangle\simeq 2.7\times 10^{11}\Msun$ \citep[assuming
Salpeter IMF;][]{Kriek08} and median effective radius
$\langle\Re\rangle=0.9 \kpc$.  These galaxies have a factor of $\sim
5$ smaller size than local ETGs of similar mass, as apparent from the
central panel of Fig.~\ref{fig:sm}, where their median location in the
$\Mstar$-$\Re$ plane can be compared with the stellar-mass--size
correlation (\ref{eq:reff}) observed locally.

We then fix $\Mstartilde$ and $\rstartilde$ such that all our seed
galaxy models have total stellar mass $\Mstar=2.7\times 10^{11}\Msun$
and $\Re=0.9 \kpc$ (symbols inside the big hexagon in
Fig.~\ref{fig:sm}).  As a consequence, the velocity unit is given, so
our seed galaxy models have projected velocity dispersions
$\sget=532\km\sminus$ ($\fstar=0.09$) and $\sget=579\km\sminus$
($\fstar=0.02$).  In the $\Mstar$-$\sget$ plane (bottom panel of
Fig.~\ref{fig:sm}) these models (represented by the left-most symbols
in the diagram) lie well above the local observed
correlation~(\ref{eq:sigma}).  We have little information on the
velocity dispersion of high-$z$ ETGs \citep[see][hereafter
vDKF09]{Cenarro09,Cap09,vDo09}: for 1255-0, an ETG at $z=2.186$ with
stellar mass $\Mstar\simeq 3\times 10^{11}\Msun$ (after conversion to
Salpeter IMF) and half-light radius $\Re=0.78\pm0.17\kpc$
\citep{Kriek09}, vDKF09 tentatively measured projected velocity
dispersion $510^{+165}_{-95} \km\sminus$, consistent within the
uncertainties with $\sget$ of our seed galaxy models (see
Fig.~\ref{fig:sm}, bottom panel).  Given the uncertainties on $\sget$,
also the structural parameter $\cstar$ is poorly constrained for
high-$z$ ETGs.  Our seed-galaxy models---which were constructed to lie
exactly on the lensing Mass Plane (NTB09)---have $\cstar=0.11$ for
$\fstar=0.09$ and $\cstar=0.13$ for $\fstar=0.02$ (left-most symbols
in the top panel of Fig.~\ref{fig:sm}), consistent, for
$\Mstar\sim 3\times 10^{11}\Msun$, with the correlation~(\ref{eq:mdim})
found for SLACS galaxies, and also with the highly uncertain value of
$\cstar$ estimated for the high-$z$ ETG 1255-0.

In Fig.~\ref{fig:sm} the points with $\log \Mstar/\Msun>11.6$ show the
location of the merger remnants of all steps of the hierarchies in the
$\Mstar$-$\cstar$, $\Mstar$-$\Re$ and $\Mstar$-$\sget$ planes.  For
the adopted values of the length and mass scales, the timescale to
complete one step of the merging hierarchies is in the range $0.5-1.7$
Gyr, so all hierarchies can be comfortably completed in the $\sim 10$
Gyr elapsing from $z\sim 2 $ to the present.  The location of the
merger remnants with respect to the local scaling laws is quite
dependent on the details of their merging history, but it is apparent
that all the merger remnants lie significantly below the $\Mstar-\Re$
relation and above the $\Mstar-\sget$ relation. On average, minor
mergers increase more the galaxy size than major mergers, consistent
with the results of \citet{Naab09}, but we also find that different
minor-merging events produce remnants with significantly different
$\Re$ and $\sget$ for similar $\Mstar$, even when the initial
conditions differ just by the relative phase-space distribution of the
centres of mass of the satellites and the central galaxy (see the
right-most pairs of stars in Fig.~\ref{fig:sm}).  

The behavior of our minor-merger hierarchies in the $\Mstar$-$\sget$
plane deserves some discussion. When a galaxy grows by accreting
satellites on almost parabolic orbits (as is the case for our
minor-merger steps; see NTB09), the virial velocity dispersion of the
remnant is lower than that of the progenitor {\it if the mass loss is
  negligible} \citep{Bez09,Naab09}. The fact that in some of our
minor-merger steps the remnant has slightly higher projected (and
virial) velocity dispersion than the progenitor must be ascribed to
the significant loss of luminous (see Fig.~\ref{fig:sm}) and dark (see
NTB09) matter, which is known to have the effect of increasing the
velocity dispersion \citep[see][]{Nip03}.  

In general the merger remnants lie close to the FP-like relation
$\Mstar$-$\cstar$, consistent with previous studies considering the FP
\citep{Gonz03,Nip03,Boy06} or the lensing Mass Plane (NTB09). An
exception is the minor-merging hierarchy with $\fstar=0.02$ seed
galaxy (crosses in the diagrams), in which $\cstar$ increases
significantly with stellar mass, reflecting the relatively low
$\Mstar$ and high $\sget$ due to substantial mass loss.

Altogether, considering both major and minor merger hierarchies, our
results confirm that dry mergers bring compact ETGs closer to the
observed correlations. However, quantitatively, the process is not
efficient enough to explain the observed size evolution.  We note that
the final merger remnants (the end-products of the last steps of the
hierarchies) have exceptionally high stellar masses ($\Mstar\gsim
10^{12}\Msun$), so one cannot invoke more growth via dry mergers to
obtain systems with surface mass density as low as observed in local
galaxies. Furthermore, velocity dispersion is approximately conserved
along the merger hierarchies, and objects with $\sget\sim
500\km\sminus$ are not observed in the local universe.

\section{Constraints on the stellar-mass assembly history of  
present-day early-type galaxies}
\label{sec:local}

Taken at face value, the stellar masses and sizes of the high-$z$ ETGs
mean that these systems cannot evolve into present-day ETGs via dry
mergers. However, possible observational biases in the high-$z$
measurements \citep[for instance, that the sizes are
underestimated;][]{Hop09b} could make the discrepancy between high and
low redshift systems smaller. Thus, if one only considers {\it
  average} properties, dry mergers may perhaps be still considered a
viable evolutionary mechanism. However, recovering the average size of
local ETGs is not sufficient: additional constraints come from the
small {\it intrinsic scatter} of observed stellar-mass scaling
relations, which constrains the dry-merger mass assembly history of
present-day ETGs, {\it independently of the actual compactness of
  their high-$z$ counterparts}.  A first attempt of constraining the
merging history of ETGs using the scatter in the stellar-mass--size
relation was performed by \citet{Shen03}, who---on the basis of an
analytic parametrization of the size growth due to dry mergers---found
that a certain degree of fine tuning in the orbital parameters of the
galaxy encounters is required in order to reproduce the local
$\Mstar$-$\Re$ relation of their SDSS sample of ETGs.  We take here a
step further, by taking advantage of N-body merger simulations and
up-to-date stellar-mass scaling laws, including information on stellar
velocity dispersion.

We quantify the growth of galaxies by dry merging by approximating the
dependence of the structural parameters on stellar mass with power-law
functions (see NTB09 for the lensing-mass analogs): $\Re\propto
\Mstar^{\betar}$, $\sget\propto\Mstar^{\betas}$ and
$\cstar\propto\Mstar^{\betac}$.  To each step of a hierarchy we
associate a value of $\betar$ by calculating $\betar\equiv \Delta \log
\Re/\Delta \log \Mstar$, where $\Delta$ indicates variation between
the remnant and the progenitor.  The resulting 30 values of $\betar$
are distributed with average $\langle\betar\rangle=1.09$ and standard
deviation $\delta\betar=0.29$. Doing the same for $\betas$ and
$\betac$, we get $\langle\betas\rangle=0.07$, $\delta\betas=0.11$ and
$\langle\betac\rangle=0.22$, $\delta\betac=0.13$.  Assuming that the
galaxy merging history is characterized by slopes $\beta$ normally
distributed with these averages and standard deviations, we can use
these quantities to describe how dry mergers move galaxies in the
planes $\Mstar$-$\cstar$, $\Mstar$-$\Re$ and $\Mstar$-$\sget$: the
relatively large values of $\delta\beta/\beta$ we find mean that dry
mergers introduce substantial scatter in these planes.

For the dry-merging scenario to be viable it must not introduce in the
scaling relations more scatter than observed.  We can use this
condition to constrain the properties of the candidate progenitors of
local ETGs, assuming as working hypothesis that they formed by dry
merging. Let us consider, for example, local ETGs with given stellar
mass $\Mstarprime$ in the range $10^{11}-10^{12}\Msun$ probed by the
SLACS sample: in Fig.~\ref{fig:scatt} the distributions (within
1-$\sigma$) of $\cstar$, $\Re$ and $\sget$ for these galaxies are
represented by the vertical bars, while the loci in the
$\Mstar$-$\cstar$, $\Mstar$-$\Re$ and $\Mstar$-$\sget$ planes of their
allowed dry-merging progenitors are shown by the shaded
areas. Galaxies lying out of these areas, growing by major or minor
dry mering, would populate regions of the parameter space inconsistent
with the local observed distributions at the reference mass
$\Mstarprime$.  It is apparent that the strongest constraint comes
from the $\Mstar$-$\Re$ plane (central panel), where the allowed
region for progenitors spans $\sim 0.26$ dex in $\Mstar$.  This means
that a growth in stellar mass by dry merging by more than a factor of
$\sim 1.8$ (and in size by more than a factor of $\sim 1.9$) is
excluded on the basis of our set of simulations. In other words,
present-day massive ($\Mstar \sim 10^{11}-10^{12}\Msun$) ETGs cannot
have assembled, on average, more than $\sim 45\%$ of their stellar
mass through dry merging.  Quantitatively, the constraints on the mass
assembly history depend on the specific choice of the orbital
parameters and mass ratios of the galaxies: in our set of merging
hierarchy simulations there are more major than minor mergers, and we
explore a wide range of orbital parameters. One might get different
numbers by considering more minor-merging simulations or excluding
parts of the orbital parameter space.  For instance, if we consider
only the 8 minor merger steps, we get $\langle\betar\rangle=1.30$ and
$\delta\betar=0.40$, which gives an upper limit of $\sim 35\%$ to the
fraction of stellar mass assembled via dry mergers by local ETGs. In
any case, it is clear that {\it a remarkable degree of fine tuning is
  required to reproduce the tightness of the local scaling relations
  with dry mergers}.

\section{Conclusions}

%As shown by our current results, as well as previous work
%\citep[e.g.,][]{Nip03,Boy06}, 

Dry merging makes galaxies less compact. For this reason it has been
proposed to be the key to reconcile the compactness of high-$z$ ETGs
with the lower stellar-mass surface-density of local ETGs.

Our quantitative analysis based on N-body simulations shows that {\it
  it is very hard to reproduce with dry mergers the growth in size of
  a factor of $\sim 5$} between $z\gsim 2$ and today reported in the
literature for $\Mstar\sim 10^{11}\Msun$ ETGs, because the average
size growth produced by dry merging is not sufficient.  The tightness
of the local scaling laws of ETGs provides an additional and stringent
constraint for the dry-merging scenario. As we have shown, generic dry
mergers tend to increase the scatter rather than decrease it because
they cannot preserve simultaneously the correlations between stellar
mass, velocity dispersion and effective radius. On the basis of our
set of major and minor merging simulations, we conclude
that---independent of the actual compactness of higher-$z$ ETGs---{\it
  typical present-day massive ETGs did not assemble more than $\sim
  45\%$ of their stellar mass and grew more than a factor $\sim 1.9$
  in size via dry merging}. A similar upper limit is found also for
the total (dark plus luminous) mass assembly on the basis of the
lensing-mass scaling laws (NTB09).

%Future samples of high-$z$ strong lenses will be needed to
%test the evolution of the lensing scaling relations.

%In conclusion, the relatively small scatter associated to the
%correlations between size (or velocity dispersion) and stellar-mass
%(or lensing mass) of local ETGs is hard to reconcile with the large
%scatter introduced in the same quantities by dry mergers.  
%The sizes and velocity dispersions of the end-products of dry-merging
%hierarchies are quite sensitive to the details of their merger
%histories.  Thus in a pure dry-merging scenario, 

Is this then the end of dry mergers as a major evolutionary mechanism?
The only way out could be provided by a high degree of fine tuning of
the mix of progenitors and orbits. Although not a natural scenario in
our opinion, this fine tuning could perhaps reproduce such tight
scaling relations and thus save the day.  A systematic exploration of
cosmologically motivated merging hierarchies is necessary to find out
whether such an extreme fine tuning actually occurs.

The question of what drives the apparent size evolution of ETGs
remains open. A possibility is that a combination of different
astrophysical processes and observational biases can do the job
\citep[e.g.][]{Hop09c}, but it is not excluded that the answer is to
be found exploring more exotic scenarios, such as those assuming
non-standard dark matter \citep{Lee08,Fer09}.

\acknowledgments

T.T. and M.W.A. acknowledge support from the NSF through CAREER award
NSF-0642621, by the Sloan Foundation through a Sloan Research
Fellowship, and by the Packard Foundation through a Packard
Fellowship, and by NASA through grants from the Space Telescope
Science Institute associated with programs \#10494, \#10798, \#11202.
Some numerical simulations were run at CINECA, Bologna, with CPU time
assigned under the INAF-CINECA agreement 2008-2010.

\clearpage

\end{document}